\documentclass[twocolumn, showpacs,prl]{revtex4}
\usepackage{graphicx}
\usepackage{dcolumn}	
\usepackage{bm}	

\begin{document}

\title{Controlled switching of intrinsic localized modes in a 1-D antiferromagnet}

\author{J. P. Wrubel}
\author{M. Sato}
\author{A. J. Sievers}
\affiliation{
Laboratory of Atomic and Solid State Physics\\
Cornell University, Ithaca, New York 14853-2501
}

\date{\today}

\begin{abstract}
Nearly steady-state locked intrinsic localized modes (ILMs) in the quasi-1d antiferromagnet (C$_{2}$H$_{5}$NH$_{3}$)$_{2}$CuCl$_{4}$ are detected via four-wave mixing emission or the uniform mode absorption. Exploiting the long-time stability of these locked ILMs, repeatable 
nonlinear switching is observed by varying the sample temperature, and localized modes with various amplitudes are created by modulation of the microwave driver power. This steady-state ILM locking 
technique could be used to produce energy localization in other atomic lattices.
\end{abstract}


\maketitle

\newpage
Whereas plane waves characterize the natural excitations of a harmonic lattice, intrinsic localized modes (ILMs) are an important feature associated with large amplitude excitations in anharmonic lattices \cite{kiselev,flach}. In arrays of Josephson-junctions \cite{binder,trias} and in a 2D photonic crystal lattice \cite{fleischer}, ILMs are produced by directly exciting their macroscopic eigenvectors. A different production method involving the modulational instability (MI) of the large amplitude uniform mode \cite{dauxois,sandusky,rossler} has been successfully used to produce ILMs in macroscopic \cite{schuster,sato_micro} and microscopic lattices \cite{sato_nature,sato_prb}. After the instability generates many localized modes, a few become stabilized by locking to a continuous wave (cw) driver frequency. These locked ILMs may be maintained indefinitely. Emission steps observed by four-wave mixing involving such locked ILMs in an antiferromagnet is providing a very sensitive method with which to examine dynamical energy localization in an atomic lattice \cite{sato_nature,sato_prb}.

Switching is a general characteristic of nonlinear systems, and in particular, localized E\&M mode switching is expected to provide an efficient method for all-optical routing in communications networks \cite{christo}. Dynamical switching is also characteristic of other kinds of physical systems such as individual driven Josephson junctions \cite{siddiqi}, and the discrete resonant breathers in Josephson junction ladders \cite{fistul}; as well as biological systems \cite{qian}. Hysteresis and switching have yet to be demonstrated with ILMs in an atomic lattice.

In this Letter we demonstrate controlled switching in locked nearly steady-state ILMs produced in the quasi-1D antiferromagnet (C$_{2}$H$_{5}$NH$_{3}$)$_{2}$CuCl$_{4}$. This is accomplished by modulating the frequency gap $\Delta f$ between the local mode and the uniform mode. Firstly, single ILM control is demonstrated via the temperature dependence of the uniform resonance. Secondly, ILMs with varying amplitudes are investigated by power modulation of the locking driver. Precise control of the numbers and amplitudes of ILMs are prerequisites for studying and manipulating these nonlinear excitations.

The experimental setup and notation we use are based on those of previous experiments with locked ILMs in the quasi-1d antiferromagnet \cite{sato_nature,sato_prb}. The AFMR frequency is sample shape dependent \cite{chikamatsu,keffer} and for c-axis directed rods occurs in the range 1.375-1.385 GHz. In earlier experiments a short, intense microwave pulse $f_{1} ~1.290$ GHz, induced the MI and produced a broad distribution of ILMs. A second microwave source with 1000 times lower power $f_{2} ~1.320$ GHz could then lock a few ILMs with nearby frequencies. These ILMs were detected by four wave mixing of $f_{2} $ with a third lower power probe $f_{3} $.  $P_{ILM}^{(3)} $, the resulting power emitted by the ILMs alone, was detected at the spectrum analyzer frequency of $f_{sp} =2f_{2} -f_{3} $. According to Eq. (10) in Ref. \cite{sato_prb}, $P_{ILM}^{(3)} $ is a function of the integer number of locked ILMs $n_{ILM} $,

\begin{equation}
\sqrt{P_{ILM}^{(3)} } =n_{ILM} f_{sp} \chi \left( f_{sp} \right) P_{2} \sqrt{P_{3} }, \label{poweq}
\end{equation}
where $\chi (f_{sp} )$ is an effective nonlinear susceptibility, and the powers delivered to the sample by $f_{2} $ and $f_{3} $ are $P_{2} $ and $P_{3} $. Because $n_{ILM}$ has integer values, this equation describes steps in the square root of the emitted power. A limitation of this method of producing locked ILMs is that emission steps have only been detected as ILMs become unlocked from the driver.

\begin{figure}[tb]
\includegraphics{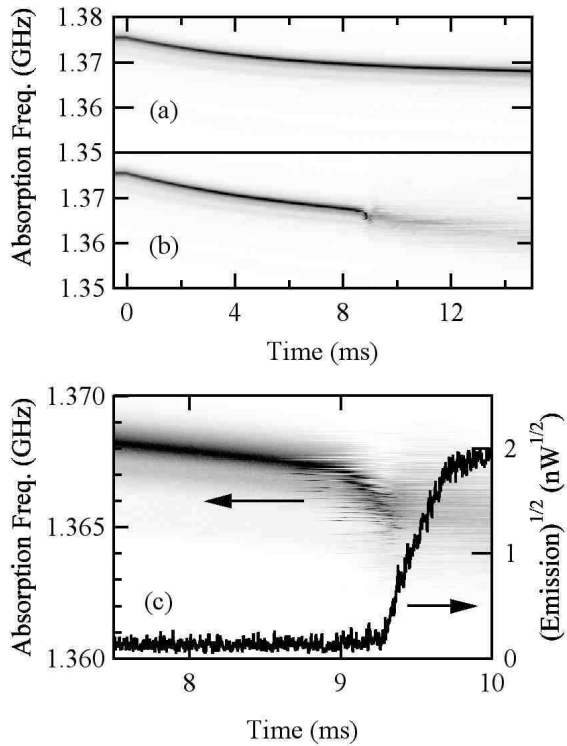}
\caption{\label{afmr} AFMR absorption versus time in the presence of a low frequency driver. $f_{2} =1.330$ GHz. Darker density represents stronger absorption. (a) Driver power = 1.1 W and (b) Driver power = 1.4 W. At about 9 ms a transition occurs to a broadened resonance. (c) Enhanced view of the broadening transition at 9 ms in (b). The four wave mixing (emission)$^{1/2}$ is superimposed. The nonlinear emission step occurs in tandem with the broadening of the AFMR.}
\end{figure}

By eliminating the short pump pulse $f_{1} $ and placing the cw driver $f_{2} $ closer in frequency to the uniform mode we are able to produce individual locked ILMs over long times such that their resulting emission steps may be observed as they form. The effect of this experimental approach is shown in Fig. \ref{afmr}. Here $f_{2} =1.330$ GHz is switched on at 0 ms. At below critical driver powers, such as shown in Fig. \ref{afmr}(a), the AFMR is pulled to slightly lower frequencies due to the soft nonlinearity of the AFMR and the increased population of finite wave number spin waves. Despite an increased spin temperature, the resonance retains its narrow width at all times. The AFMR eventually reaches a nearly steady-state frequency as the energy input by the driver is balanced by relaxation to the lattice. In Figure \ref{afmr}(b) the driver power is sufficient for energy localization to occur. At 9 ms, the AFMR is rapidly pulled down; and shortly afterwards it reforms at nearly the same frequency, but significantly broadened. The broadening is accompanied by a step increase in the nonlinear emission. The detailed AFMR transition from narrow to broad is expanded in Fig. \ref{afmr}(c). The traces were acquired individually for each frequency and were not averaged. The differences from frequency to frequency are indicative of the uncertainty in the energy localization time. A typical time dependent trace of the square root of the emitted ILM power is shown for reference.

\begin{figure}[tb]
\includegraphics{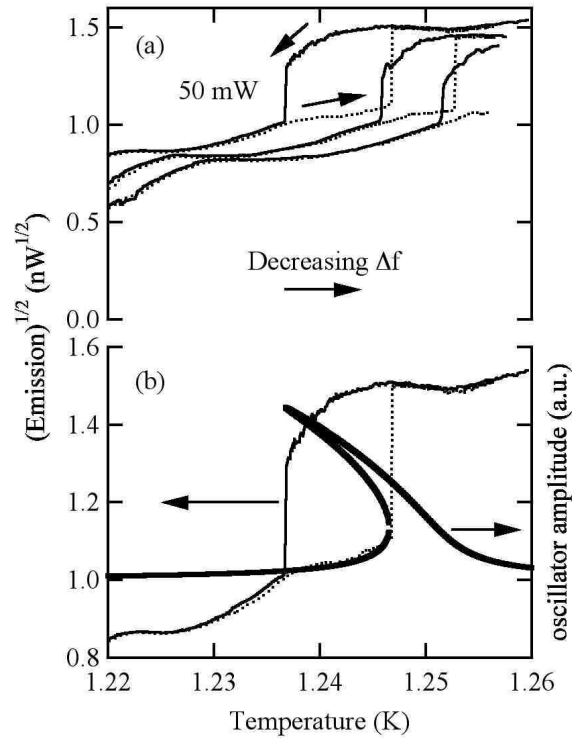}
\caption{\label{temps} Step hysteresis in (emission)$^{1/2}$ versus temperature. (a) $f_{2} =1.350$ GHz with three powers: 50 mW, 47.4 mW, and 45.7 mW. Dotted lines: increasing temperatures; solid lines: decreasing temperatures. The hysteresis due to capture and loss of a single ILM is evident. (b) Comparison of the 50 mW data with a model. Thick line: hysteresis characteristic of the amplitude response for a driven nonlinear oscillator. The abscissa is the temperature dependence of the AFMR in the presence of the driver $f_{2}$.}
\end{figure}

The long lifetime of the localized mode produced with this locking technique allows us to probe the step behavior by varying the sample temperature.  In this way a very gentle manipulation of the local mode gap $\Delta f$ may be accomplished since the AFMR frequency varies with temperature. In Fig. \ref{temps}(a) the measured (emission)$^{1/2}$ from a steady-state locked ILM is shown as a function of temperature, for three different driver powers. The largest cw power is 50 mW. Sample temperature rates of no more than a few mK sec$^{-1}$ are used. As the temperature increases, the AFMR moves towards $f_{2} $. At 1.247 K, the emission takes a step up as an ILM becomes locked. At higher temperatures the emission is nearly constant. On subsequently decreasing the temperature, the emission is flat until about 1.237 K where there is a roll-off, an ILM is then lost, and the emission steps back to its original level. The step size (emission)$^{1/2}$ of 0.4 nW$^{1/2}$ (0.3 nW$^{1/2}$ ) with increasing (decreasing) temperature is quite similar with ILM step sizes observed previously \cite{sato_nature,sato_prb}. There is an interval of 10 mK between the two ILM switching temperatures. At lower $f_{2} $ powers the steps move to higher temperatures and the hysteresis width decreases slightly. This is the first observation of reversible switching behavior of an intrinsic localized mode in an atomic crystal.

The shape of the switching hysteresis resembles that expected from the response of a driven nonlinear oscillator near its fundamental frequency \cite{jordan}.  Here the amplitude of the nonlinear oscillator corresponds to the 3rd order nonlinear AC magnetization, which is proportional to the square root of the emitted power, as shown in Ref. \cite{sato_prb} (Eq. 3). The nonlinear oscillator amplitude is plotted in Fig. \ref{temps}(b) as a function of its resonant frequency, which varies as the measured temperature dependent AFMR. For this nearly classical 1-D spin system the temperature dependence of the magnetization is given by a Langevin function, and is approximately linear well below the transition temperature \cite{morrish}. Although the hysteresis matches very well, there are clear differences; nonetheless, the nonlinear oscillator model suggests that the ILM hysteresis may be understood qualitatively within this simple picture.

\begin{figure}[tb]
\includegraphics{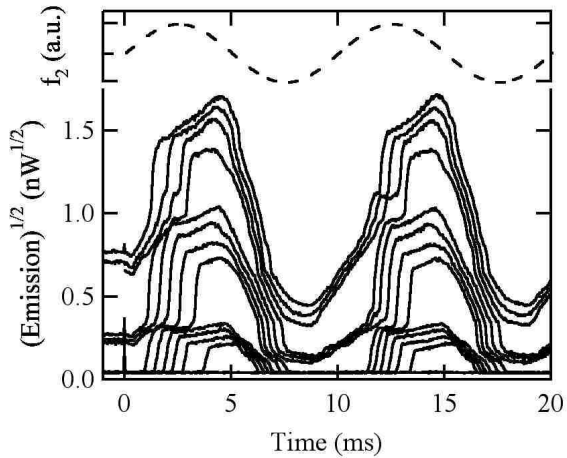}
\caption{\label{pow1} (Emission)$^{1/2}$ versus time during $f_{2} $ power modulation. Solid lines plot the ILM emission as $f_{2} $ is modulated at 100 Hz and scanned from 112 to 200 mW in 4.7 \% increments from bottom to top. The 25 \% sine wave amplitude modulation begins at $t=0$ ms. The dashed curve illustrates the time pattern of $f_{2}$.  The emission has a phase shift of $~ -\pi/5$ relative to the driver.}
\end{figure}

The second way of controlling the local mode gap $\Delta f$ is by modulating the power of the locking driver.  With a sine wave amplitude modulation, as the power increases, the AFMR is pulled closer to $f_{2} $ and ILMs become locked; then as the power decreases, the gap increases and ILMs are unlocked. Figure \ref{pow1} shows an example from this type of experiment with $f_{2} =1.330$ GHz and a 100 Hz sine wave modulation with 25 \% amplitude. The solid traces from bottom to top represent a subset of the acquired data with increasing driver powers. The (emission)$^{1/2}$ has a complex stepped structure with the same period as the applied field, and is reproducible over multiple modulation periods.

\begin{figure}[tb]
\includegraphics{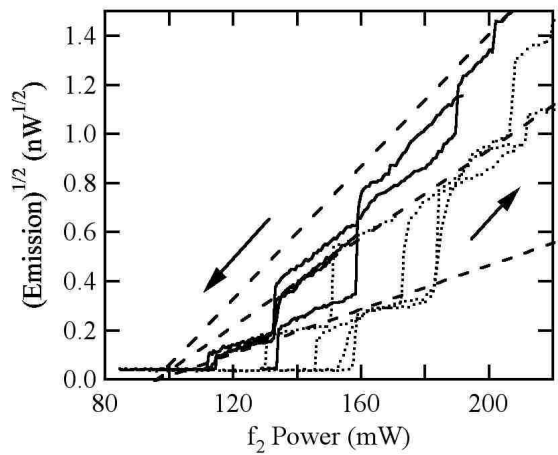}
\caption{\label{pow2} (Emission)$^{1/2}$ at constant times as a function of $f_{2} $ power. The traces extracted from Fig. \ref{pow1} are separated by 1.25 ms. Dotted data lines: $10-15$ ms (ILM creation); Solid data lines: $15-20$ ms (ILM loss). The $f_{2} $ power on the abscissa is calculated separately for each slice, and includes a phase shift of $-\pi /5$ to that of the applied field. The dashed lines are guides to the eye that follow the proportionality given in Eq. (\ref{poweq}).}
\end{figure}

The complexity shown in Fig. \ref{pow1} may be reduced by plotting the power dependence of (emission)$^{1/2}$ at a few constant time slices. This is done in Fig. \ref{pow2}. Slices are shown at 1.25 ms intervals during the second modulation period from 10 - 20 ms. The abscissa of each trace corresponds to the instantaneous applied power at that time, but with the phase shift deduced from Fig. \ref{pow1} included. The step height and switching power depend significantly on the emission slope at which the slice was made. The dotted lines in Fig. \ref{pow2} are representative of times from 10 - 15 ms during which ILMs are becoming locked to the driver. These increasing emission traces have larger steps and occur at higher $f_{2} $ powers.  The solid lines are representative of times from 15 - 20 ms, during which ILMs are becoming unlocked from the driver. These decreasing emission traces have smaller steps and occur at lower $f_{2} $ powers.  Modulation dependent hysteresis is apparent; with ILMs becoming locked with increasing emission on the right edge, and then becoming lost with decreasing emission on the left edge. The width of the hysteresis loop depends on the rate of change in the emission signal - the more rapidly the change, the wider it is.

The dashed lines in Fig. \ref{pow2} suggest that the square root of the emission step height is proportional to the $f_{2}$ power. The lines have slopes of 0.0045 nW$^{1/2}$ mW$^{-1}$ per ILM and x-intercepts at 95 mW. The first two steps fit very well, whereas there is significant variation in the third step. The linear dependence of $\sqrt{P_{ILM}^{(3)} } $ on $P_{2}$ for each individual ILM is correctly predicted by Eq. (\ref{poweq}), and differs from the lack of power dependence found previously \cite{sato_nature,sato_prb}. This difference can be traced to the importance of the local mode gap $\Delta f$ in determining the ILM amplitude and size. The earlier experiments did not detect any power dependence because ILMs with identical amplitude were sequentially lost from the locking driver. In contrast, the power dependence in Fig. \ref{pow2} suggests that ILMs with different amplitudes are observed in the different traces. The larger steps imply narrow ILMs and a larger gap, while the smaller steps imply wide ILMs with smaller gaps. 

A question yet to be addressed is how an ILM actually forms in this steady state experiment. Figure \ref{afmr}(c) shows that the width of the AFMR absorption line remains narrow as its frequency drops. Somewhat later the ILM forms as seen by the four wave emission signal. This is not the signature of the MI. Instead we propose that at the spatial location of the large AFMR spin amplitude, perhaps due to sample inhomogeneity, the frequency is pulled more than at the small amplitude locations. This increases the coupling to $f_{2} $ and drives additional amplitude into that localized region. When its frequency reaches that of $f_{2} $ a locked ILM is formed. With the addition of each ILM the AFMR linewidth increases in an incremental fashion. It is puzzling that despite the small size of the ILM \cite{sato_nature}, it has such a dramatic effect on the width of the AFMR.

We have shown that the switching behavior of steady-state locked ILMs in the quasi-1d antiferromagnet (C$_{2}$H$_{5}$NH$_{3}$)$_{2}$CuCl$_{4}$ can be controlled by varying the frequency gap between the ILM and the AFMR. The appearance and disappearance of ILMs can be monitored both by four-wave mixing emission and AFMR absorption. Temperature modulation of the local mode gap yielded a method for switching a locked spin ILM. A steady-state ILM perturbed by power modulation is found to switch with a linear dependence of step height on driving power, demonstrating a controllable variation in its amplitude.

These spin wave experiments suggest an analogous method for generating ILMs in other systems such as an anharmonic phonon system. By temperature tuning a paraelectric crystalline film \cite{comes} so that its TO mode is close in frequency to that of a high power THz laser line, the ILM generation procedure described here, which does not rely on modulational instability ignition, can be carried out. Now a degenerate four wave mixing output \cite{shen} from the same fixed frequency laser could be used to identify atomic ILMs.

\begin{acknowledgments}
We thank J. B. Page for fruitful discussions. JPW would like to acknowledge the U.S. Department of Education for a graduate fellowship. This work was supported by NSF-DMR Grant No. 0301035 and by the Department of Energy Grant No. DE-FG02-04ER46154.
\end{acknowledgments}


\begin{thebibliography}{99}
\bibitem{kiselev} S. A. Kiselev, S. R. Bickham, and A. J. Sievers, Comm. in Cond. Mat. Phys. \textbf{17}, 135 (1995).
\bibitem{flach} S. Flach and C. R. Willis, Phys. Repts. \textbf{295}, 182 (1998).
\bibitem{binder} P. Binder, D. Abraimov, A. V. Ustinov, S. Flach, and Y. Zolotaryuk, Phys. Rev. Lett. \textbf{84}, 745 (2000).
\bibitem{trias} E. Tr\'{\i}as, J. J. Mazo, and T. P. Orlando, Phys. Rev. Lett. \textbf{84}, 741 (2000).
\bibitem{fleischer} J. W. Fleischer, M. Segev, N. K. Efremidis, and D. N. Christodoulides, Nature \textbf{422}, 147 (2003).
\bibitem{dauxois} T. Dauxois and M. Peyrard, Phys. Rev. Lett. \textbf{70}, 3935 (1993).
\bibitem{sandusky} K. W. Sandusky and J. B. Page, Phys. Rev. B, \textbf{50}, 866 (1994).
\bibitem{rossler} T. R\"{o}ssler and J. B. Page, Phys. Rev. B \textbf{62}, 11460 (2000).
\bibitem{schuster} M. Schuster, F. Pignatelli, and A. V. Ustinov, Phys. Rev. 
B \textbf{69}, 094507 (2004).
\bibitem{sato_micro} M. Sato, B. E. Hubbard, A. J. Sievers, B. Ilic, D. A. Czaplewski, and H. G. Craighead, Phys. Rev. Lett. \textbf{90}, 044102 (2003).
\bibitem{sato_nature} M. Sato and A. J. Sievers, Nature \textbf{432}, 486 (2004).
\bibitem{sato_prb} M. Sato and A. J. Sievers, Phys. Rev. B \textbf{71}, 214306 (2005).
\bibitem{christo} D. N. Christodoulides, F. Lederer, and Y. Silberberg, Nature \textbf{424}, 817 (2003).
\bibitem{siddiqi} I. Siddiqi \textit{et al.}, Phys. Rev. Lett. \textbf{94}, 027005 (2005).
\bibitem{fistul} M. V. Fistul, A. E. Miroshnichenko, S. Flach, M. Schuster, and A. V. Ustinov, Phys. Rev. B \textbf{65}, 174524 (2002).
\bibitem{qian} H. Qian and T. C. Reluga, Phys. Rev. Lett. \textbf{94}, 028101 (2005).
\bibitem{chikamatsu} M. Chikamatsu, M. Tanaka, and H. Yamazaki, J. Phys. Soc. Jpn \textbf{50}, 2876 (1981).
\bibitem{keffer} F. Keffer, in \textit{Handbuch der Physik}, edited by S. Fl{\underline{ue}}gge (Springer-Verlag, Berlin, 1966), Vol. XVIII/2, 
\bibitem{jordan} D. W. Jordan and P. Smith, \textit{Nonlinear Ordinary Differential Equations}, Third ed. (Oxford University Press, Oxford, 1999).
\bibitem{morrish} A. H. Morrish, \textit{The Physical Principles of Magnetism} (John Wiley \& Sons, Inc., New York, 1965).
\bibitem{comes} R. Comes and G. Shirane, Phys. Rev. B \textbf{5}, 1886 (1972).
\bibitem{shen} Y. R. Shen, \textit{The Principles of Nonlinear Optics} (John Wiley \& Sons, New York, 1984), p. 249.
\end{thebibliography}
\end{document}